%% file: singletop_taujets_D0.tex
\documentclass[aps,prl,twocolumn,showpacs,floatfix]{revtex4}

\usepackage{graphics}
\usepackage{color}
\usepackage{fancybox}
\usepackage{setspace}
\usepackage{balance} 
\usepackage[british]{babel} 
\usepackage{hyphenat}
\usepackage{graphicx}
\usepackage{amssymb}
\usepackage{amsbsy}
\usepackage{amsmath}
\usepackage{slashed}
\usepackage{enumerate}
\usepackage{subfigure}
\usepackage{hyperref}

\newcommand{\TauXSobs} {$3.4^{+2.0}_{-1.8}$}

\newcommand{\TauXSobsPeakWidth} {1.9}
\newcommand{\TauXSexpPeakWidth} {1.8}

\newcommand{\SuperCombineXSexpPeakWidth} {4.7}

\newcommand{\EMUCombineXSexpPeakWidth} {4.5}

\newcommand{\SuperCombineExpGainInPeakWidth} {4.4\%}

\newcommand{\TauLimObs} {7.3}

\newcommand{\ttbarXSobs} {$10.0^{+2.3}_{-1.6}$~pb}

\newcommand{\tauola}    {\sc{tauola}}
\newcommand{\SuperCombineXSall} {3.84^{+0.89}_{-0.83}}
\newcommand{\met} {$\slashed{E}_T$}
\newcommand{\dzero} {D0}
\newcommand{\ttbar} {$t\bar{t}$}

\begin{document}
%\shadowbox{
%\begin{minipage}{2.5in}
%\begin{minipage}{6.7in}
%{\color{red} INTERNAL DOCUMENT (Version \version)} to be submitted to
\begin{tabular}{p{13cm}r}
&FERMILAB-PUB-09-610-E\\
\end{tabular}

\title{\textbf{Search for single top quarks in the tau+jets channel using 4.8~fb$^{-1}$ of $p\bar{p}$ collision data}}
\input{list_of_authors_r2.tex}
\begin{abstract}
%% Text of abstract
  We present the first direct search for single top quark production
  {\color{black}using reconstructed tau leptons in the final state}.
  The search is based on 4.8~fb$^{-1}$ of integrated luminosity
  collected in $p\bar{p}$ collisions at $\sqrt{s}$=1.96~TeV with the
  D0 detector at the Fermilab Tevatron Collider. We select events with
  a final state including an isolated tau lepton, missing transverse
  energy, two or three jets, one or two of them being identified as
  $b$ quark jet. We use a multivariate technique to discriminate
  signal from background. The number of events observed in data in
  this final state is consistent with the signal plus background
  expectation. We set in the tau+jets channel an upper limit on the
  single top quark cross section of \TauLimObs~pb at the $95\%$~C.L.
  This measurement allows a gain of 4\% in expected sensitivity for
  the observation of single top production when combining it with
  electron+jets and muon+jets channels already published by the D0
  collaboration with 2.3~fb$^{-1}$ of data. We measure a combined
  cross section of $\SuperCombineXSall$~pb, which is the most precise
  measurement to date.
\end{abstract}
\pacs{14.65.Ha; 14.60.Fg; 13.85.Rm}
\maketitle
% \begin{keyword}
% %% keywords here, in the form: keyword \sep keyword
% %% PACS codes here, in the form: \PACS code \sep code
% %% MSC codes here, in the form: \MSC code \sep code
% %% or \MSC[2008] code \sep code (2000 is the default)
% Top quark production\sep tau lepton\sep electroweak interaction\sep multivariate analysis\\
% \PACS 14.65.Ha\sep 14.60.Fg\sep 12.15.-y\sep 13.85.Rm\sep 02.50.Sk
% % http://www.aip.org/pacs/pacs08/pacs08-toc.html 14.65.Ha Top quarks
% % 14.60.Fg Taus 12.15.-y Electroweak interactions 13.85.Rm Limits on
% % production of particles 02.50.Sk Multivariate analysis
% \end{keyword}
%\modulolinenumbers[2]
%\setpagewiselinenumbers
%\linenumbers
%%%%%%%%%%%%%%%%%%%%%%%
%              _       
%  /\/\   __ _(_)_ __  
% /    \ / _` | | '_ \ 
%/ /\/\ \ (_| | | | | |
%\/    \/\__,_|_|_| |_|
%----------------------
%\lefthyphenmin=5
%\righthyphenmin=5
\section{Introduction}
\begin{figure}[h]
\centering
\includegraphics[width=0.5\textwidth]{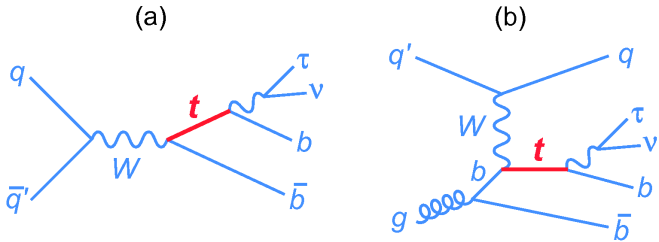}
\caption{Diagrams for single top quark production in (a)
  $s$-channel and (b) $t$-channel production showing the top quark
  decays of interest.}\label{fig:tbtqb}
\end{figure}

At the Fermilab Tevatron Collider, top quarks can be produced either
in pairs by the strong interaction or singly by the electroweak
interaction. Single top quark production can be used to directly
measure the CKM matrix element $|V_{tb}|$~\cite{paper.1992.Jik}, to
determine the top quark partial decay width and
lifetime~\cite{1995car}, to study top quark polarization and to probe
physics beyond the standard model (SM)~\cite{paper.2000.Tai.STtheory}.
The production of a single top quark is accompanied by a $b$ quark in
the $s$-channel mode or by both a $b$ quark and a light quark in the
$t$-channel mode, as illustrated in Fig.~\ref{fig:tbtqb}.
Besides the $s$- and $t$-channels, single top quarks can also be
produced in an associated $tW$ process via $bg\rightarrow tW$. At the
Tevatron, this channel has a negligible cross section compared to $s$-
and $t$-channel production~\cite{st-theory}. The $s$-channel process
is referred to as ``$tb$'' production, where $tb$ includes both
$t\bar{b}$ and $\bar{t}b$ states.  The $t$-channel process is
abbreviated as ``$tqb$'', where this includes $tq\bar{b}$,
$t\bar{q}\bar{b}$, $\bar{t}qb$, and $\bar{t}\bar{q}b$ states.
Considering the SM decay modes of the top quark and $W$ boson, single
top production and decay results in four channels: electron+jets,
muon+jets, tau lepton+jets (tau+jets), and all-jets.
Evidence~\cite{evidence-09fb-prl,evidence-09fb-prd,cdf_st_prl2008} and
observation~\cite{d0_st_prl2009,cdf_st_prl2009} of single top quark
production in the electron+jets and muon+jets channels and the first
direct measurement of $|V_{tb}|$~\cite{evidence-09fb-prl} have been
published recently. However, the tau+jets channel has not been
measured so far due to the overwhelming jet background at the
Tevatron, although signatures involving tau leptons have been explored
by {\dzero}, for example, in the measurement of the
$Z\rightarrow\tau\tau$ cross section~\cite{paper.2005.D0.ZTauTau} and,
more recently, in the context of Higgs searches~\cite{paper.higgstau}.
The measurement of the single top quark cross section in the electron
and muon channels is still limited by statistical uncertainties. In
this analysis, adding the tau+jets channel increases the signal
acceptance by 32\% compared to the {\dzero}
observation~\cite{d0_st_prl2009}. In addition, the tau+jets channel is
a statistically independent channel with different dominant
backgrounds and different systematic contributions compared to the
electron and muon channels. As such, the tau channel provides an
independent measurement of the single top production cross section. In
addition, the approach developed in the analysis could be extended to
other studies, such as Higgs searches in tau+jets channels, where the
cross section to be measured is low and multijet background is
dominant.

In this Letter, we report the first direct search for single top
quarks in the tau+jets channel. Since the dominant background source
is multijet events, which are poorly modeled by Monte Carlo (MC)
simulations, we build most of our background model from an independent
sample of multijet data. We then model several smaller background
sources using MC and combine them with the multijet sample to complete
the background model. We then train a multivariate discriminant to
separate the simulated single top signal from the background model.
Finally, we extract the single top cross section and combine the
result with the existing electron and muon channel measurements.

\section{Object Identification and Event Selection}\label{sect.id}
The upgraded D0 detector is described in detail in
Ref.~\cite{paper.2006.Aba-d0upgrade}. A right-handed coordinate system
is used in the analysis. In the system, the $z$-axis is along the
proton direction, $\phi$ is the azimuthal angle, $\eta$ is the
pseudorapidity, $-\ln{[\tan{\theta/2}]}$, where $\theta$ is the polar
angle, and the true rapidity is defined as,
$1/2\ln{[(E+p_zc)/(E-p_zc)]}$~\cite{paper.2006.Aba-d0upgrade}.  This
analysis is based on a sample of D0 Run~II data collected between
August 2002 and April 2009.  Run~IIa and Run~IIb data are defined as
two sub-datasets corresponding to integrated luminosities of
1.0~fb$^{-1}$ and 3.8~fb$^{-1}$ respectively. A new inner layer of
silicon microstrip tracking detectors was added to the detector
between Run~IIa and Run~IIb. The additional tracking detectors and the
increased instantaneous luminosity in Run~IIb change the $b$-jet
identification performance.

The sample considered contains events which have passed one of a
list of specialized trigger conditions. The most important ones either
set a threshold on the total scalar sum of transverse momenta of the
jets in the event, require a minimum transverse momentum of all jets,
or select events based on the acoplanarity of the two leading jets
sorted in transverse energy. The trigger efficiency in this analysis
is $\approx$45\%.

A hadronically decaying tau lepton appears as a narrow jet in the
{\dzero} detector. A tau candidate is a calorimeter cluster
reconstructed from all the towers with energy above a threshold within
a cone $\mathcal{R}\equiv\sqrt{(\Delta\phi)^2+(\Delta\eta)^2}<$0.5
(where $\phi$ is the azimuthal angle and $\eta$ is the pseudorapidity)
around a seed tower.  The tau candidate must have at least one track
associated with the cluster, and possibly an additional energetic
subcluster of cells in the electromagnetic (EM) section of the
calorimeter~\cite{paper.2005.D0.ZTauTau}. Hadronic tau candidates are
separated in three types according to the tracking and EM calorimeter
information: (1) single track with no EM subclusters, (2) single track
with EM subclusters, and (3) two or three associated tracks. The
classification is motivated by the decay modes: (1) $\tau^\pm
\rightarrow \pi^\pm \nu $ (2) $\tau^\pm \rightarrow \rho ^\pm \nu $
and (3) $\tau^\pm \rightarrow \pi^\pm \pi^\pm \pi^\mp (\pi^0)\nu $. We
require the tau transverse momentum, $p_T^{\tau}$, to be larger than
10, 5, 10~GeV for Type 1, 2 and 3 tau leptons.  We also require the
transverse momentum of the associated track, $p_T^{\mathrm{trk}}$, to
be larger than 7~GeV (5~GeV) for Type 1~(2) tau leptons.  For Type 3,
the transverse momentum of at least one track, $p_T^{\mathrm{trk}}$,
has to be larger than 5~GeV and the sum of the associated track
transverse momenta, $\sum_{\mathrm{trk}}p_T^{\mathrm{trk}}$, has to be
greater than 7~GeV.  Hadronically decaying tau leptons are
distinguished from other types of jets using variables such as
isolation (defined as $(E^\tau_T -
E_T^{\mathrm{core}})/E^\mathrm{core}_T$ where $E^\tau_T$ and
$E^\mathrm{core}_T$ are the transverse energy in a cone with radius
$\mathcal{R}=0.5$ and a smaller cone with radius $\mathcal{R}=0.3$
defined about the same axis), shower width, and shower profiles (a
ratio of the $E_T$'s of the two most energetic calorimeter towers with
size $\Delta\phi\times\Delta\eta=0.1\times 0.1$ over $E_T$ of the tau
candidate).  As there is no single tau identification variable which
can provide the required background rejection, a multivariate
technique is used to combine these features into a single
discriminant.  Tau identification is then performed by applying
kinematic selections as well as a requirement on the multivariate
discriminant output.

All other analyses at D0 that use hadronic tau decays except the one
reported in this Letter use tau identification relying on a neural
network (NN) trained on $Z\rightarrow\tau\tau$ decays and background
samples suitable for that signal~\cite{paper.2005.D0.ZTauTau}. In
contrast, the multivariate technique used in this analysis for tau
identification relies on boosted decision trees (BDT). The BDT
technique has been used in previous D0 single top quark
analyses~\cite{evidence-09fb-prl,d0_st_prl2009} and is described in
Ref.~\cite{evidence-09fb-prd}. In brief, a decision tree is an
algorithm which combines selection requirements on a large number of
variables with varying discriminating power into a single, more
powerful, multivariate
discriminant~\cite{paper.1984.Bre,paper.1993.Bow}. It can be
``boosted'' by building the multivariate discriminant through a
weighted average score from many decision trees instead of a single
tree~\cite{paper.1996.Fre}. {\color{black}A total of 25 well-modeled}
kinematic variables {\color{black}for each tau type} serve as the
inputs to BDTs.  {\color{black}Table~\ref{tbl.tauvar} shows the 10
  most discriminative variables with their normalized importance
  values for tau Types 1, 2 and 3. The importance is derived by an
  algorithm in which variable usage frequency, separation gains and
  numbers of events in the splitting nodes are
  considered~\cite{varimportance}.}  A set of trees is created based
on a simulated tau sample from single top quark MC events, and
realistic background strongly dominated by fake tau leptons. This fake
tau background is extracted from data by requiring events to pass tau
jet triggers and applying the kinematic selections given above.  Both
the signal and background have different kinematics from the standard
NN training samples.  By changing the technique from neural networks
to boosted decision trees we gain $\approx$3\%, $\approx$8\% and
$\approx$2\% (for Types 1, 2 and 3) signal efficiency for the same
background rejection rate (98\%). By changing both the technique and
the signal and background samples to match the busy single top
environment with extra jets we gain $\approx$8\%, $\approx$20\% and
$\approx$8\% (for Types 1, 2 and 3) signal efficiency yielding
$\approx$76\%, $\approx$69\% and $\approx$59\% for the same rejection.
We require exactly one tau lepton per event.

\begin{center}
\begin{table*}[hbt]\color{black}
  \caption{The 10 most discriminative variables with their normalized importance values in the training of the tau identification BDT. The variables listed are explained in Appendix~I.}\label{tbl.tauvar}
\begin{tabular}{p{0.2cm}l|p{0.1cm}p{2.5cm}p{2.2cm}|p{0.1cm}p{2.5cm}p{2.2cm}|p{0.1cm}p{3.cm}p{2.cm}}\hline\hline
&Rank\ \ && Tau Type 1 & Importance && Tau Type 2 & Importance && Tau Type 3 & Importance   \\ \hline
&1 && Width$_{\eta,\phi}(\tau)$    &  2.1$\times 10^{-1}$ &&        Width$_{\eta,\phi}(\tau)$  &  2.8$\times 10^{-1}$ &&                                Isolation  &  7.1$\times 10^{-1}$ \\
&2 &&                Isolation     &  2.0$\times 10^{-1}$ &&                          Profile  &  2.8$\times 10^{-1}$ &&             Ratio$_{\tau,\mathrm{trks}}$  &  3.9$\times 10^{-2}$ \\
&3 &&Ratio$_{\tau,\mathrm{trks}}$  &  9.1$\times 10^{-2}$ &&              $e_{12}$  &  1.5$\times 10^{-1}$ &&                      $e_{12}$  &  3.0$\times 10^{-2}$ \\
&4 &&$E^\mathrm{trk1}_T$/$E^\tau_T$&  6.9$\times 10^{-2}$ &&                 Profile$^\prime$  &  5.5$\times 10^{-2}$ &&$E_T^\mathrm{alltrk-trk1-trk2}$/$E^\tau_T$ &  2.5$\times 10^{-2}$ \\
&5 &&            Profile$^\prime$  &  5.1$\times 10^{-2}$ &&                   $\delta\alpha$  &  2.3$\times 10^{-2}$ &&       $z^\mathrm{trk1}_\mathrm{DCA}$      &  1.9$\times 10^{-2}$ \\
&6 &&$E_T^\mathrm{EMlayer3}$/$E^\tau_T$  &  3.5$\times 10^{-2}$ &&   Profile$_\mathrm{layer3}$ &  2.3$\times 10^{-2}$ &&                Width$_{\eta,\phi}(\tau)$  &  1.6$\times 10^{-2}$ \\
&7 &&          Isolation$^\prime$  &  3.2$\times 10^{-2}$ &&     Ratio$_{\tau,\mathrm{trks}}$  &  2.0$\times 10^{-2}$ &&           $E^\mathrm{trk1}_T$/$E^\tau_T$  &  1.5$\times 10^{-2}$ \\
&8 &&                     Profile  &  3.2$\times 10^{-2}$ &&  $E_T^\mathrm{EMcl2}$/$E^\tau_T$  &  1.7$\times 10^{-2}$ &&                 Profile$_\mathrm{layer3}$ &  1.3$\times 10^{-2}$ \\
&9 &&  Ratio$_\mathrm{EM12,\tau}$  &  3.0$\times 10^{-2}$ &&   $E^\mathrm{trk1}_T$/$E^\tau_T$  &  1.7$\times 10^{-2}$ &&         Width$^\prime_{\eta,\phi}(\tau)$  &  1.1$\times 10^{-2}$ \\
&10 &&         Width$^\prime_{\eta,\phi}(\tau)$  &  2.8$\times 10^{-2}$ &&          Isolation  &  1.5$\times 10^{-2}$ &&          $E_T^\mathrm{EMcls}$/$E^\tau_T$  &  1.1$\times 10^{-2}$ \\ \hline\hline
% &1 &&                $rms$  &  2.078$\times 10^{-1}$ &&             $rms$  &  2.785$\times 10^{-1}$ &&              trkiso  &  7.095$\times 10^{-1}$ \\
% &2 &&             trkiso  &  1.955$\times 10^{-1}$ &&             profile  &  2.774$\times 10^{-1}$ &&          ET\_o\_sum  &  3.859$\times 10^{-2}$ \\
% &3 &&         ET\_o\_sum  &  9.073$\times 10^{-2}$ &&             e1e2/ET  &  1.541$\times 10^{-1}$ &&             e1e2/ET  &  2.995$\times 10^{-2}$ \\
% &4 &&            ett1/ET  &  6.863$\times 10^{-2}$ &&            profile2  &  5.459$\times 10^{-2}$ &&             ettr/ET  &  2.496$\times 10^{-2}$ \\
% &5 &&           profile2  &  5.068$\times 10^{-2}$ &&              dalpha  &  2.258$\times 10^{-2}$ &&               tzDCA  &  1.932$\times 10^{-2}$ \\
% &6 &&         EM3\_Et/ET  &  3.486$\times 10^{-2}$ &&                prf3  &  2.182$\times 10^{-2}$ &&                 $rms$  &  1.642$\times 10^{-2}$ \\
% &7 &&               iso2  &  3.196$\times 10^{-2}$ &&          ET\_o\_sum  &  1.969$\times 10^{-2}$ &&             ett1/ET  &  1.493$\times 10^{-2}$ \\
% &8 &&            profile  &  3.190$\times 10^{-2}$ &&        emcl\_et2/ET  &  1.723$\times 10^{-2}$ &&                prf3  &  1.276$\times 10^{-2}$ \\
% &9 &&           EM12isof  &  3.005$\times 10^{-2}$ &&             ett1/ET  &  1.715$\times 10^{-2}$ &&                $rms2$  &  1.106$\times 10^{-2}$ \\
% &10 &&               $rms2$  &  2.836$\times 10^{-2}$ &&              trkiso  &  1.473$\times 10^{-2}$ &&             empt/ET  &  1.093$\times 10^{-2}$ \\ \hline\hline
\end{tabular}
\end{table*}
\end{center}

Jets are reconstructed by an iterative cone algorithm with radius
$\mathcal{R}=0.5$ in rapidity-azimuth space~\cite{report.jetid}. The
highest-$p_T$ jet must have $p_T>25$~GeV and the second highest-$p_T$
jet $p_T>20$~GeV while any additional jet must have $p_T>15$~GeV.  The
highest-$p_T$ jet must have pseudorapidity $|\eta|<$~2.5 and all other
jets $|\eta|<$~3.4.  The jets and the tau lepton must be isolated by
requiring for their spatial separation in pseudorapidity-azimuth space
be larger than 0.5.  In order to identify $b$ jets, a neural network
is trained on the outputs of three $b$-jet identification algorithms:
secondary vertex, jet lifetime probability, and counting signed impact
parameter~\cite{paper.2009.bid-nim}. All three of these algorithms
discriminate $b$ jets from light quark jets by exploiting the
signatures of the relatively long lifetime of $b$ hadrons. If the
neural network output of a jet is larger than 0.775, the jet is tagged
as a $b$ jet. This operating point corresponds in our selected sample
to a $b$-tagging efficiency of 40\% and a light-quark tagging rate of
0.4\%. We select events with two or three jets, including at least one
$b$ jet, in order to enhance the signal-to-background ratio.  We also
require $20<$\met$<200$~GeV where \met\ is the missing transverse
energy which is equal to the negative of the vectorial sum of the
transverse energy deposited in the calorimeter by all particles. A tau
energy scale correction has been applied and \met\ has been corrected
for the presence of the tau leptons. We do not exclude electrons that
satisfy the tau identification requirement since these electron events
provide $>$50\% of our signal acceptance.  However, we veto events
with one isolated electron or one isolated muon to make sure the
tau+jets sample has no overlap with the electron and muon samples in
order to be able to combine the measurements. The data have been split
by tau (Types 1 and 2 combined and Type 3), jet multiplicity (two jets
and three jets), number of $b$ jets (one $b$ jet and two $b$ jets) and
running period, for a total of 16 analysis channels.

We select 3845 $b$-tagged tau+jets candidate events, among which we
expect 72 single top quark events. Table~\ref{tbl.yield} shows the
event yields for all channels combined. About 85\% of single top quark
events in this sample come from tau Types 1 and 2 and 86\%
are events with only one $b$ jet. The acceptance times efficiency is
3.0\% when considering only hadronic tau leptons.

\begin{table}[h]
\centering
\caption{Expected and observed events in 4.8 fb$^{-1}$ of integrated luminosity shown in tau Types 1 and 2, Type 3 channels and all analysis combined. The uncertainties include both statistical and systematic components. }\label{tbl.yield}
\begin{tabular}{p{2.5cm}r@{ $\pm$ }lr@{ $\pm$ }lr@{ $\pm$ }l} \hline\hline
  Source   &  \multicolumn{2}{c}{Types 1 and 2}  & \multicolumn{2}{c}{Type 3} & \multicolumn{2}{c}{Sum}  \\ \hline
  $tb$+$tqb$ & 61&11 & 11 & 2  &          72 & 12  \\ 
  $W$+jets   & 573&68 & 107 & 12 & 680 & 104    \\
  $Z$+jets   & 43&8 & 17 & 3 & 60 & 10     \\
  Dibosons   & 30&5 & 7 & 1 & 37 & 6      \\
  $t\bar{t}$ & 170&35&60 &12&  230 & 44     \\
  Multijets  & 1444&38 &1182 &21&  2626 & 98    \\
  %Total prediction &  2260 & 93 &1373 &28 & 3633 & 153 \\ 
Total prediction &  2321&94 &1384 &28 & 3705 & 153 \\ 
  Data &\multicolumn{2}{c}{\mbox{2372\ \ }} &\multicolumn{2}{c}{\mbox{\ \ 1473}} &\multicolumn{2}{c}{\mbox{\ 3845}}\\
\hline\hline
\end{tabular}
\end{table}

\section{Signal and Background Modeling}\label{sect.model}
Single top quark events are simulated by the next-to-leading order
(NLO) event generator {\sc singletop}~\cite{paper.2006.comphep},
which is based on {\sc comphep}~\cite{paper.1999.Puk,paper.2004.Boo}.

Since tau leptons are observed as narrow jets of particles in the
calorimeter, the main background to single top quark events in the
tau+jets channel is multijet production. This is unlike the other
leptonic single top channels in which $W$+jets events are the main
background~\cite{d0_st_prl2009,cdf_st_prl2009}. We have developed a
method to model the multijet background directly from data. The
principal steps in this method can be summarized as:
\begin{enumerate}
\item Derive a tag rate function (TRF) to describe the probability to
  $b$-tag any individual jet in the sample.
\item Apply this TRF to the data sample that has no $b$-tagged jets.
 \item Using simulated events for other physics sources, subtract them
   from the sample derived in Step 2 to get ``pure-multijets''.
 \item Normalize the sample derived in Step 3 to data.
 \item Combine the derived background sample, pure-multijets, with
   simulations of other background sources: $t\bar{t}$, $W$+jets,
   $Z$+jets, dibosons.
\end{enumerate}

In Step 1, we take the ratio of the number of $b$-tagged jets in our
data sample to the total number of jets to define a tag rate: the
average probability that a jet is identified as a $b$ jet. We measure
the tag rate as a function of jet $p_T$ and $\eta$ and jet
multiplicity.

In Step 2, we apply these TRFs to those events that have no $b$-tagged
jets. This TRFed sample is kinematically similar to our analysis
sample, but there is no overlap since we require at least one
$b$-tagged jet in our analysis sample.

In Step 3, we remove physics background sources such as $t\bar{t}$,
$W$+jets, $Z$+jets and dibosons. In this procedure, we subtract from
the zero-tagged TRFed multijet sample the contaminations of
$t\bar{t}$, $W$+jets, $Z$+jets and dibosons. Other background sources
are modeled through simulations. These simulations, except the tau
decay, have been described in~\cite{d0_st_prl2009}. The program
{\tauola}~\cite{tauola} (version 2.5) was used to model the decays of
tau leptons including polarization effects. We normalize the $W$+jets
background to match data by the scale factors that are derived from
the study in the electron+jets and muon+jets
channels~\cite{d0_st_prl2009}. We apply TRFs to the zero-tagged MC
samples to estimate the contamination mentioned above. A similar
procedure is used to ensure that any small single top signal
contamination in the background data sample is also subtracted.

In Step 4, the multijet events after contamination removal are
normalized to data in a multijets-enriched region, as defined by the
background-dominated region of the multivariate discriminant described
below.

In addition to multijet events modeled by the procedure described
above, our background model includes $t\bar{t}$, $W$+jets, $Z$+jets
and dibosons modeled directly from simulation. In Step 5, we
combine these simulated samples with the data-derived multijet
sample.

\begin{figure*}[hbt]
\centering
\subfigure[]{
\includegraphics[width=0.45\textwidth]{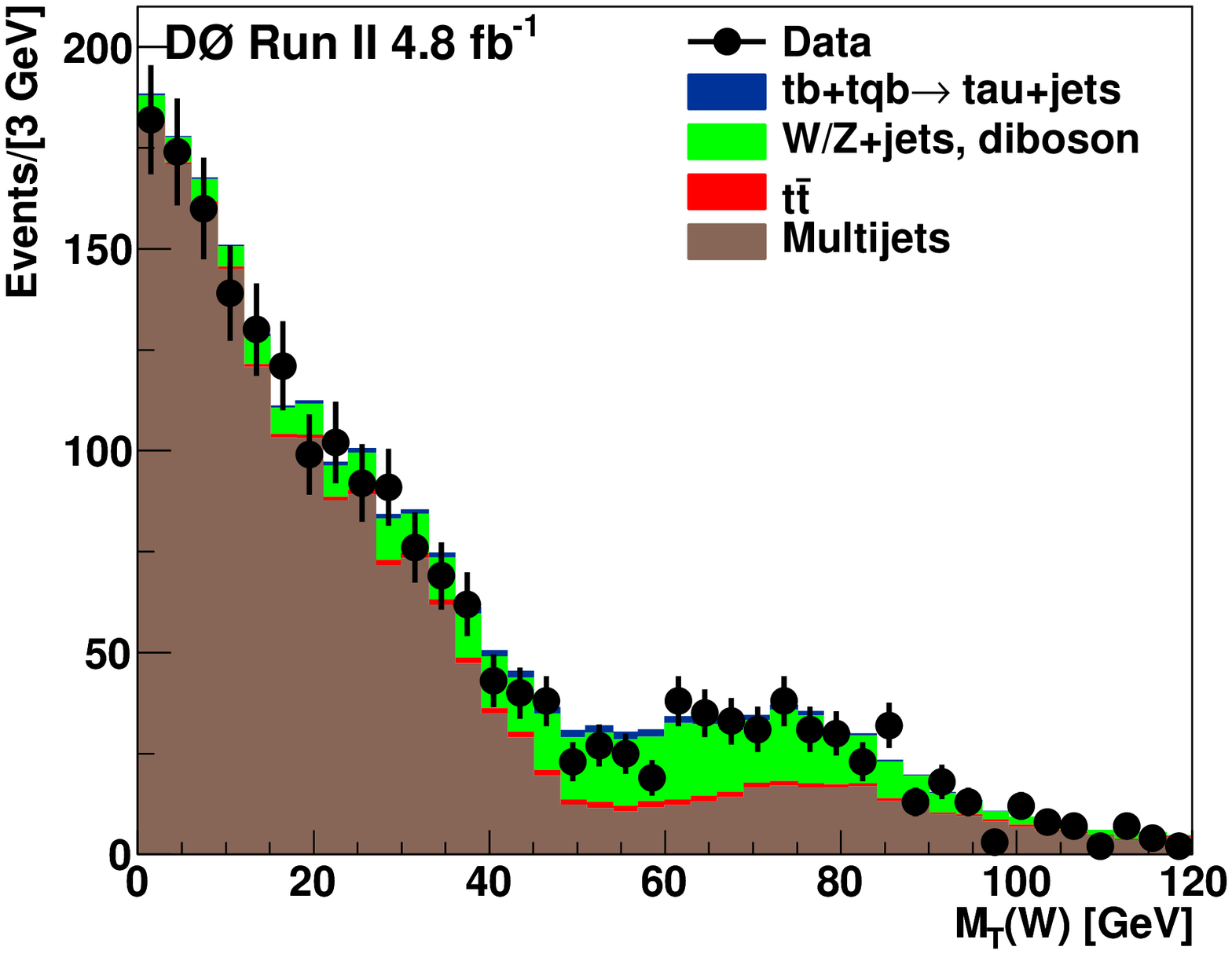}}
\subfigure[]{
\includegraphics[width=0.45\textwidth]{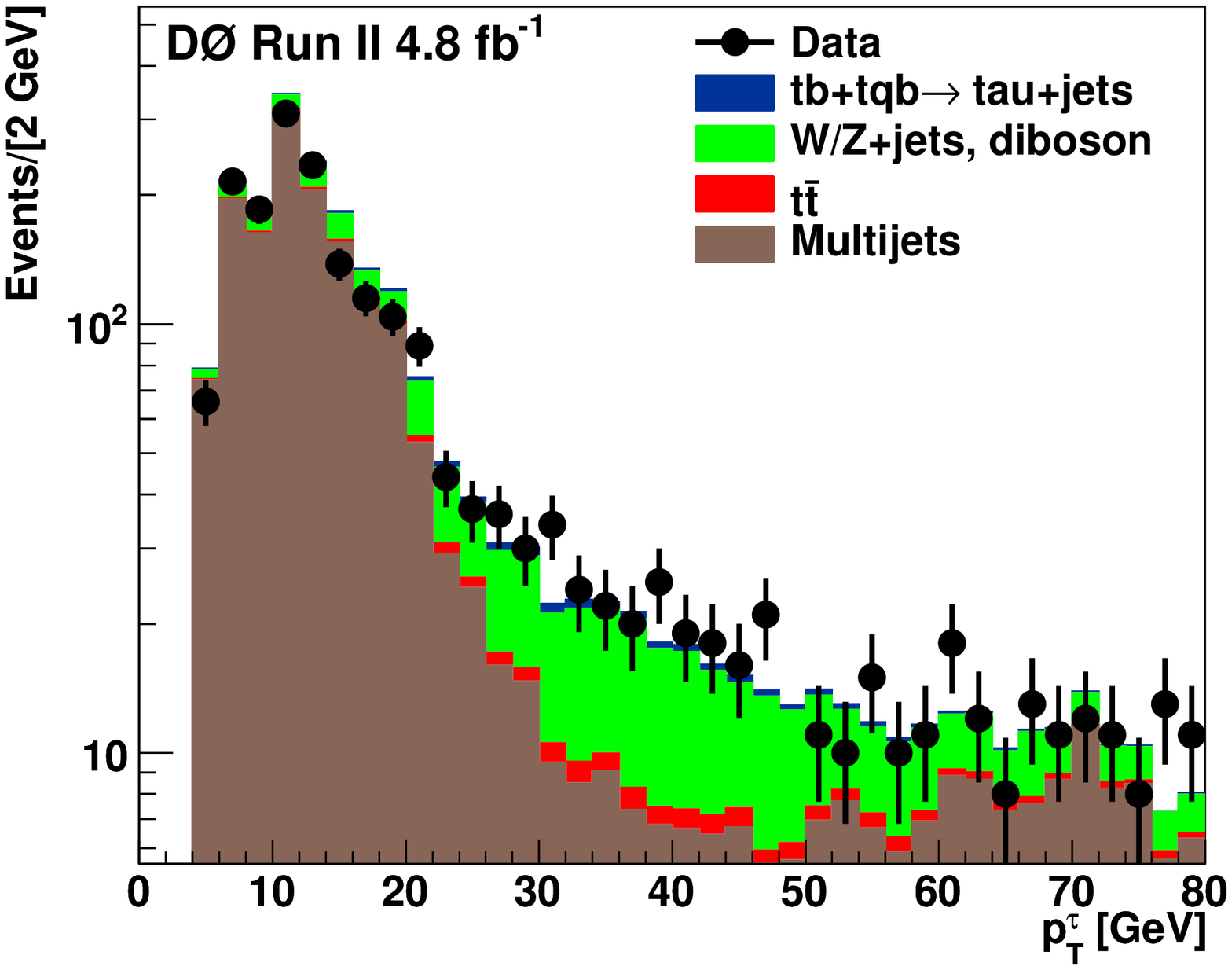}}
\subfigure[]{
\includegraphics[width=0.45\textwidth]{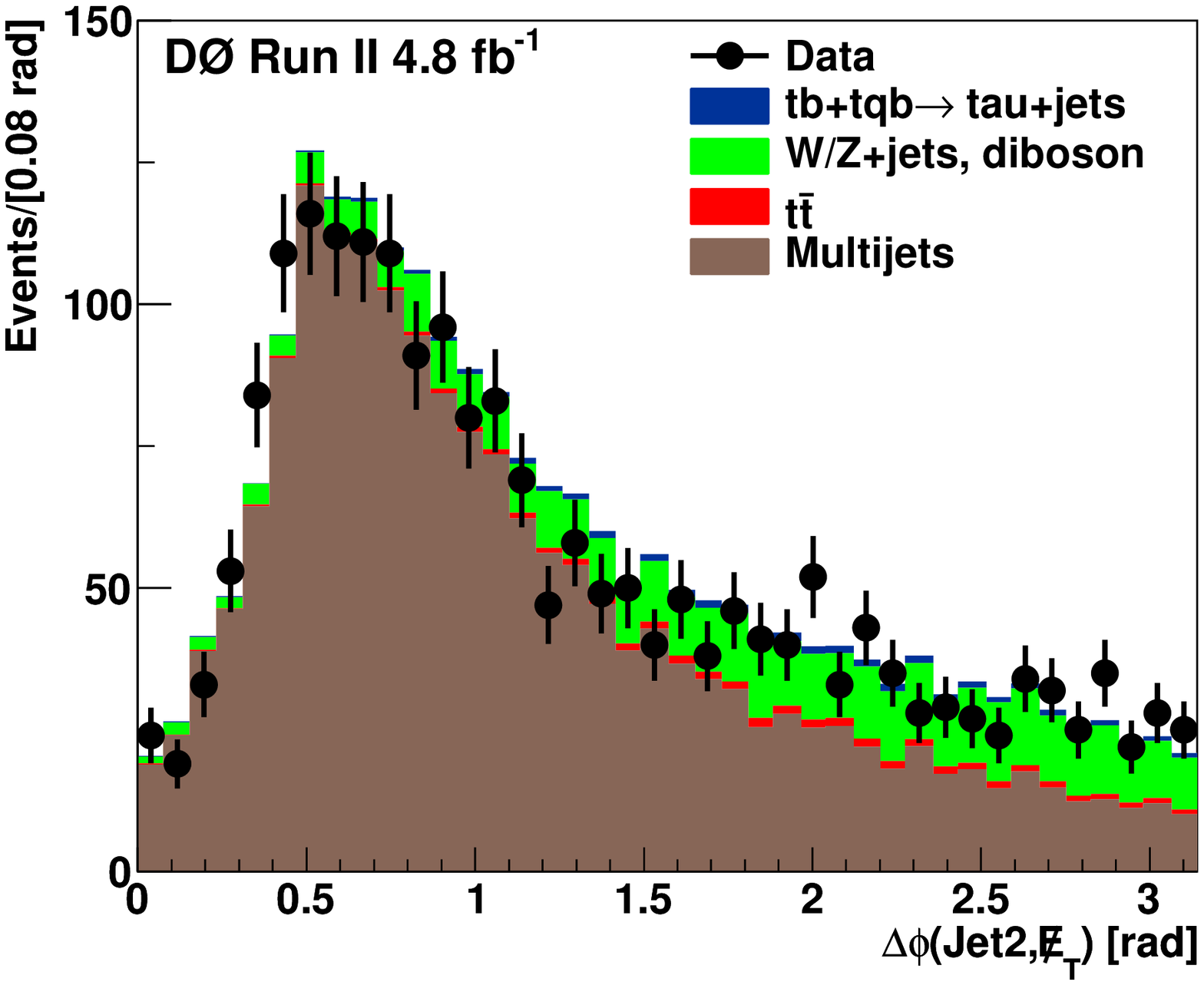}
}
\subfigure[]{
\includegraphics[width=0.45\textwidth]{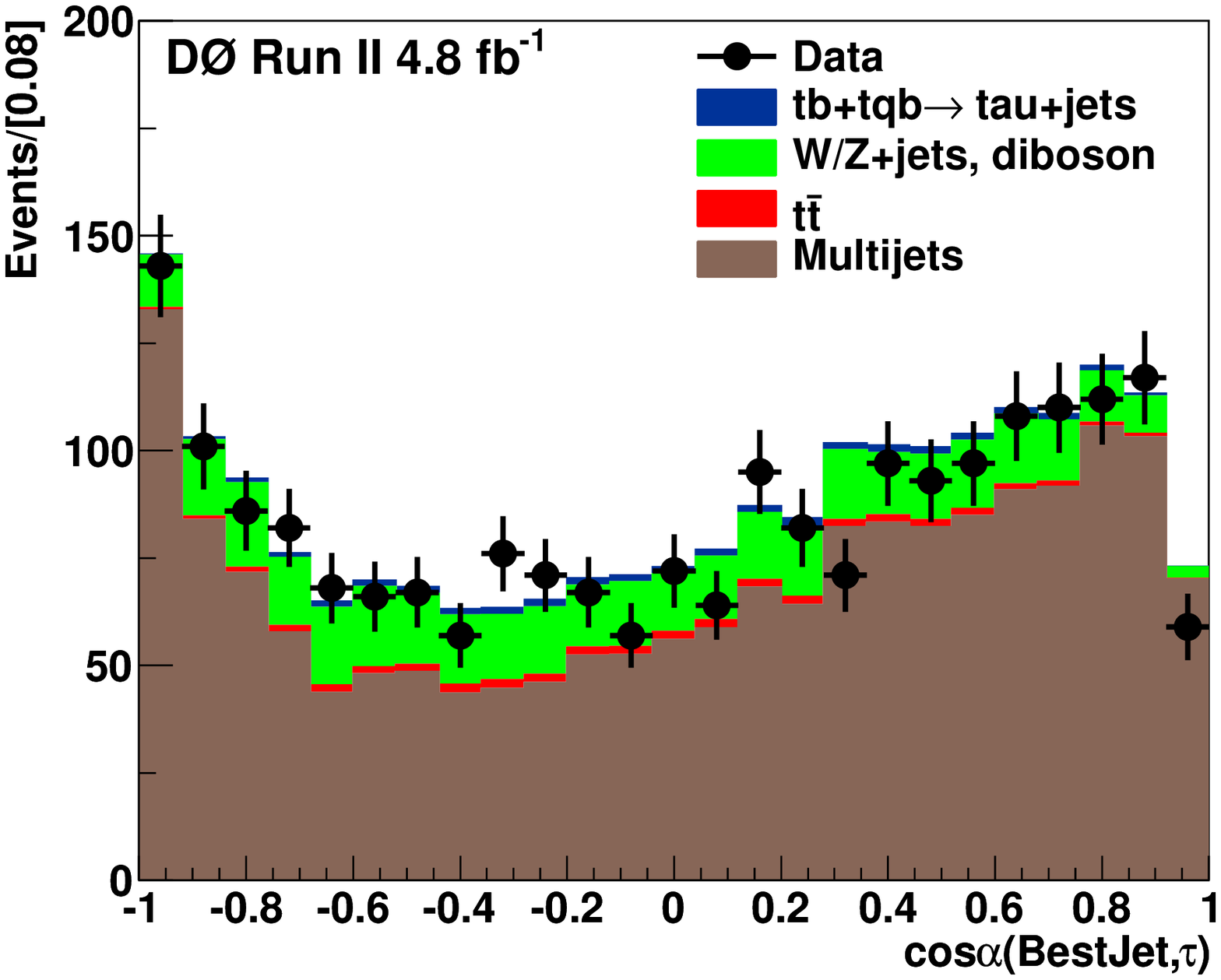}
}
\caption{Comparison between data and background distributions in the
  most sensitive channel: two jets, one $b$ tag, tau Types 1 and 2
  combined.  (a) $W$ boson transverse mass (b) tau transverse momentum
  (c) azimuthal angle between the second-highest-$p_T$ jet and
  $\slashed{E}_T$, and (d) cosine of the angle between the tau and {\color{black}the
  best jet candidate} that is used to reconstruct the best top quark
  mass (defined as closest to 170~GeV). In (b), the double-peak
  structure is caused by different $p_T$ threshold for tau types 1 and
  2.}\label{fig.vars}
\end{figure*}

At the end of the background modeling procedure, we investigate
approximately 150 topological variables to confirm that data and the
background model are in good agreement since it is expected that the
single top quark events represent only a small fraction, $\approx$2\%,
of the selected data sample. The variables can be categorized in four
classes: object kinematics, jet reconstruction, top quark
reconstruction and angular correlations.  Figure~\ref{fig.vars} shows
four discriminating variables: $W$ boson transverse mass, tau
transverse momentum, azimuthal angle between the second-highest-$p_T$
jet and $\slashed{E}_T$, and cosine of the angle between the tau
lepton and a jet candidate that is used to reconstruct the best top
quark mass (defined as closest to 170~GeV).  These variables are shown
for the most sensitive channel: Types 1 and 2, two jets, one of them
$b$ tagged.

\section{Boosted Decision Trees}\label{sect.bdts}
It is expected that single top quark events are only a small fraction
of the selected data sample.  We use the BDT technique to separate the
signal from the background.  We also employ the Kolmogorov-Smirnov
(KS) test~\cite{link.ks} to verify the compatibility of discriminating
variables in shape between data and background model.  From the
$\approx$150 variables studied, 44 to 70 of them are selected as input
variables to train BDTs depending on the individual analysis channels.
We select only variables which have a KS probability $>0.1$.  The KS
values of the selected variables are uniformly distributed above this
value.

\begin{table}[h]
  \caption[var]{\color{black}The 15 most discriminative
    BDT training variables with their normalized importance values in the most
    sensitive channel. $\Delta\phi(\mathrm{obj1},\ \mathrm{obj2})$ is the azimuthal angle between obj1 and obj2. $\cos\alpha(\mathrm{obj1,\ obj2})$ is cosine of the angle between obj1 and obj2. ``jet1'' and ``jet2'' are the highest-$p_T$ jet and the second-highest-$p_T$ jet, respectively. ``jet1+jet2'' is a system consisting of ``jet1'' and ``jet2''. The subscript, ``TopFrame'', indicates that the reference frame is the rest frame of a top quark which is reconstructed using a $b$-tagged jet, while the subscript ``tag'' (``untag'') refers to the jet passing (failing) the $b$-jet identification algorithm. $\sum_\mathrm{trks\ in\ evt} p_T^\mathrm{trk}$ is the transverse momentum of the vectorial sum of all tracks with a cut on the distance of closest approach (DCA) to the primary vertex. }\label{tbl.var}
\begin{center}\small\color{black}
\begin{tabular}{lp{5cm}c}\hline\hline
  Rank\ \ \ & \ \ \ \ \ \ \ \ \ \ \ \ Variable & Importance \\ \hline
  1 & $W$ boson transverse mass             & 5.0$\times 10^{-1}$ \\
  2 & $\Delta\phi$(jet2, $\slashed{E}_T$)   & 3.0$\times 10^{-1}$ \\
  3 & $p_T(\tau)$              & 3.3$\times 10^{-2}$ \\
  4 & $\cos\alpha$(best jet, $\tau$)           & 2.8$\times 10^{-2}$ \\
  5 & $p_T$(jet1+jet2)                      & 1.8$\times 10^{-2}$ \\
  6 & $\Delta\phi$($\tau$, $\slashed{E}_T$) & 1.5$\times 10^{-2}$ \\
  7 & $\cos\alpha$(tag, $\tau$)$_\mathrm{TopFrame}$       & 1.1$\times 10^{-2}$ \\
  8 & $\Delta\phi$(jet1, $\slashed{E}_T$)                       & 1.1$\times 10^{-2}$ \\
  9 & $\sum_\mathrm{trks\ in\ evt} p_T^\mathrm{trk}$  & 1.0$\times 10^{-2}$ \\
  10 & Best top quark mass                   & 1.0$\times 10^{-2}$ \\
  11 & $p_T(\mathrm{best\ jet})$          & 7.6$\times 10^{-3}$ \\
  12 & $Q(\tau)\times\eta$(untag) & 5.6$\times 10^{-3}$ \\
%13 & $z_\mathrm{primary vertex}$          & 4.928$\times 10^{-3}$ \\
%14 & $\mathcal{R}$($\tau$, Jet2)             & 4.633$\times 10^{-3}$ \\
%15 & $\mathcal{R}$($\tau$, Jet1)             & 3.897$\times 10^{-3}$ \\ \hline\hline
13 & $z$ position of primary vertex          & 4.9$\times 10^{-3}$ \\
14 & $\mathcal{R}(\tau$, jet2) & 4.6$\times 10^{-3}$ \\
15 & $\mathcal{R}(\tau$, jet1) & 3.9$\times 10^{-3}$ \\ \hline\hline
%   Rank\ \ \ & \ \ \ \ \ \ \ \ \ \ \ \ Variable & Importance \\ \hline
%   1 & $W$ boson transverse mass             & 4.967$\times 10^{-1}$ \\
%   2 & $\Delta\Phi$(jet2, $\slashed{E}_T$)   & 2.973$\times 10^{-1}$ \\
%   3 & $p_T^\tau$              & 3.298$\times 10^{-2}$ \\
%   4 & $\cos\phi$(best jet, $\tau$)           & 2.811$\times 10^{-2}$ \\
%   5 & $p_T$(jet1+jet2)                      & 1.771$\times 10^{-2}$ \\
%   6 & $\Delta\Phi$($\tau$, $\slashed{E}_T$) & 1.509$\times 10^{-2}$ \\
%   7 & $\cos\phi$(tag, $\tau$)$_\mathrm{btaggedtop}$       & 1.126$\times 10^{-2}$ \\
%   8 & $\Delta\Phi$(jet1, $\slashed{E}_T$)                       & 1.121$\times 10^{-2}$ \\
%   9 & $\sum_\mathrm{trks\ in\ evt} p_T^\mathrm{trk}$  & 1.030$\times 10^{-2}$ \\
%   10 & Best top quark mass                   & 1.003$\times 10^{-2}$ \\
%   11 & $p_T^\mathrm{best\ jet}$          & 7.562$\times 10^{-3}$ \\
%   12 & $Q(\tau)\times\eta$(untag) & 5.552$\times 10^{-3}$ \\
% %13 & $z_\mathrm{primary vertex}$          & 4.928$\times 10^{-3}$ \\
% %14 & $\mathcal{R}$($\tau$, Jet2)             & 4.633$\times 10^{-3}$ \\
% %15 & $\mathcal{R}$($\tau$, Jet1)             & 3.897$\times 10^{-3}$ \\ \hline\hline
% 13 & $z$ position of primary vertex          & 4.928$\times 10^{-3}$ \\
% 14 & Spatial separation between $\tau$ and jet2 & 4.633$\times 10^{-3}$ \\
% 15 & Spatial separation between $\tau$ and jet1 & 3.897$\times 10^{-3}$ \\ \hline\hline
\end{tabular}
\end{center}
\end{table}

\begin{figure}[h]
\centering
\includegraphics[width=0.45\textwidth]{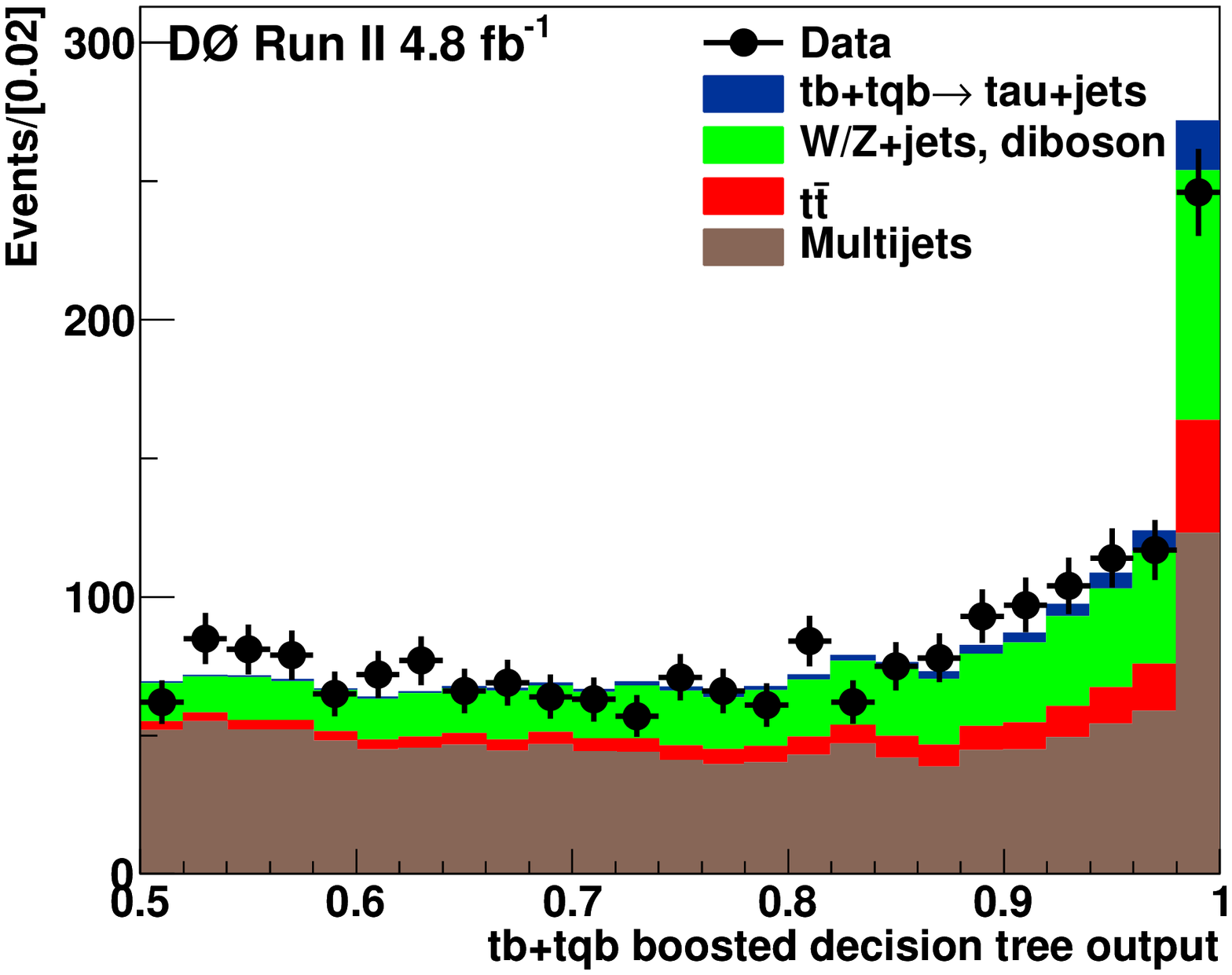}
\caption{Distribution of BDT output with all channels combined in the
  signal region (BDT$>$0.5). The single top quark signal
  ($tb+tqb\rightarrow$tau+jets) is normalized to the measured cross
  section.}\label{fig.bdt}
\end{figure}

Separate sets of BDTs are built with these variables for each analysis
channel. {\color{black}Table~\ref{tbl.var} lists the 15 most
  discriminative variables with their normalized importance
  values in the most sensitive channel. } Figure~\ref{fig.bdt} shows
the BDT output with all channels combined, in the region between 0.5
to 1.0, i.e. where the single top quark signal events are expected.
Data and the background model are in good agreement in the region. The
background-dominated region from 0.0-0.2 is used to define the
multijets-enhanced region used in Step 4 of the multijet background
modeling procedure.

\section{Systematic Uncertainties}\label{sect.error}
We consider systematic uncertainties from correction factors applied when modeling the signal and
background~\cite{evidence-09fb-prd}. ``Normalization'' uncertainty
components from the correction factors affect the signal efficiency
and the normalization of the
background samples, while ``shape'' uncertainties
change the shapes of the distributions for the background and the
expected signal. The largest
uncertainties arise from 
$W$+jets normalization to data, 
tau identification efficiency, 
tag rate functions, 
and jet-flavor correction in $W$+jets and $Z$+jets events.
Other uncertainties include 
multijets normalization, 
integrated luminosity,  
MC statistics, 
jet energy scale, 
jet identification, jet energy resolution, 
initial- and final-state radiation,  
jet fragmentation, 
theoretical cross sections,  
the reweighting of the jet angular distributions in $W$+jets events,  
signal contamination removal, non-multijets contamination, branching fractions, instantaneous luminosity reweighting, parton distribution functions, primary vertex selection, and tau energy scale.  
The total uncertainty on the background model is 4.2\%--19\% depending
on the analysis channel. Table~\ref{tbl.sys}
summarizes all sources of uncertainties considered.

Some of the uncertainties are common with the study in the
electron and muon channels and have been presented in
Ref.~\cite{evidence-09fb-prd}. Below are the uncertainties specific to
this analysis:
\begin{enumerate}[(i)]
\item \textit{$W$+jets normalization to data} (normalization)\\
  The uncertainty is on the scale factors applied to normalize
  $W$+jets to match data. Since we use the scale factors derived from
  the electron+jets and muon+jets study~\cite{d0_st_prl2009}, we
  consider the difference between these two channels as the
  uncertainty in the tau+jets channel.
\item \textit{Multijets normalization} (normalization)\\ The statistical
  uncertainty of the multijet sample in the BDT region [0.0, 0.2] is
  used.
\item \textit{Tag rate functions} (shape and normalization) \\ This
  uncertainty consists of two components: those on the multijet
  background sample and those on the MC samples related to $b$-tag
  modeling.  The former is evaluated by raising and lowering the tag
  rate by one standard deviation of its experimental determination.
  Uncertainties considered in the latter are from several sources:
  statistics of the simulated events; the assumed heavy flavor
  fractions in the simulated multijet sample used for the mistag rate
  determination; and the choice of
  parameterizations~\cite{evidence-09fb-prd}.
\item \textit{Tau identification efficiency} (normalization)\\
  This uncertainty is estimated by the difference in tau identification
  efficiency between data and MC as derived in a tau-enriched data
  sample.
\item \textit{Signal contamination removal} (shape and
  normalization)\\ In Step 3 of the background modeling, we reweight
  single top quark events to remove any small signal contamination.
  The uncertainty is evaluated by raising and lowering the weighting
  function by one~standard deviation.
\item \textit{Non-multijets contamination removal} (shape)\\
  In Step 3 of the background modeling, we subtract the non-multijets
  contamination from the zero-tagged TRFed multijet sample by
  weighting events. The uncertainty is evaluated by raising and
  lowering the weighting function by one~standard deviation.
\item \textit{Tau energy scale} (normalization)\\ The energy of
  hadronic tau candidates with low energy is corrected using the
  energy in the calorimeter and the momentum of the tracks associated
  to the tau leptons using parameterized single pion response
  functions. The uncertainty on the scale is estimated by varying
  these parameterizations.
\end{enumerate}

\begin{table}[h]
  \caption[gensys]{A summary of the relative systematic uncertainties for each of the correction factors or normalizations. The uncertainty shown is the relative error on the correction or the efficiency, before it has been applied to the MC or data samples. We do not show relative systematic uncertainties of the components for shape since they depend on distribution binning.}\label{tbl.sys}
\begin{center}\small
\begin{tabular}{p{6.5cm}r}\hline\hline
\multicolumn{2}{c}
{\underline{Relative Systematic Uncertainties}} \vspace{0.1in}\\
\hline
{\bf Components for Normalization}      &              \\
~~$b$-jet fragmentation                 &  2.0\%       \\
~~Branching fractions                   &  1.5\%       \\
~~Diboson cross sections                &  5.8\%       \\
~~Instantaneous luminosity reweighting  &  1.0\%       \\
~~Integrated luminosity                 &  6.1\%       \\
~~Initial- and final-state radiation   &  (0.6--8.0)\% \\
~~Jet energy resolution                 &  4.0\%       \\
~~Jet energy scale                      &  (4.0-14.0)\%       \\
~~Jet fragmentation                     &  5.0\%       \\
~~Jet identification &  1.0\%       \\
~~MC statistics                         & (0.5--16.0)\%   \\
~~Parton distribution functions         &  3.0\%       \\
~~(signal acceptances only)             &              \\
~~Primary vertex selection              &  1.4\%       \\
~~Multijets normalization                     & (3.0--14.0)\% \\
~~Tau energy scale                      &  (1.0--1.5)\% \\
~~Tau identification efficiency         &  11.0\%       \\
~~Triggers                              &  5.5\%       \\
~~{\ttbar} cross section                &  12.7\%      \\
~~$W$+jets heavy-flavor fraction        & 13.7\%         \\
~~$W$+jets normalization to data                & (7.0--15.0)\% \\
~~$Z$+jets cross section                &  3.6\%       \\
~~$Z$+jets heavy-flavor fraction        & 13.7\%         \\\hline
{\bf Components for Shape}              &              \\
~~{\sc Alpgen} reweighting on $W$+jets sample & ---  \\
~~Non-multijets contamination removal         & ---  \\ \hline
{\bf Components for Shape and Normalization}              &       \\
~~Signal contamination removal          & ---   \\
~~Tag rate functions                   & ---   \\ \hline\hline
\end{tabular}
\end{center}
\end{table}

\section{Results}\label{sect.st}
The number of events observed in data and the shape of the BDT
discriminant are consistent with the sum of the signal and background
predictions. To estimate the statistical significance of the signal
observation we use the same Bayesian approach as in
Refs.~\cite{evidence-09fb-prl,evidence-09fb-prd,d0_st_prl2009}. This
involves forming a binned likelihood as a product over all bins and
channels. When measuring a cross section, its central value is defined
by the position of the peak in the posterior density, and the 68\%
interval about the peak is taken as the uncertainty. The posterior
density is integrated from 0 until 95\% of the posterior area is
contained and the upper limit is set at this point.  Systematic
uncertainties, including all correlations, are reflected in this
posterior interval.  Assuming a single top quark cross section of
3.46~pb for a top quark mass of 170~GeV~\cite{st-theory}, we estimate
the expected sensitivity to the standard model signal by calculating
the ratio of the position of the peak of the expected posterior
density to its lower half width.  This yields a ratio of
\TauXSexpPeakWidth, i.e. a sensitivity corresponding to approximately
1.8 standard deviation.

In order to test the linearity of our procedure with respect to the
single top quark cross section, we generate several ensembles
of pseudodatasets by randomly sampling from background model events.
We specify five input signal cross sections: 2.0~pb, 3.46~pb, 6.0~pb,
8.0~pb and 10.0~pb and generate ensembles at each value.  Each
ensemble contains $\approx$2000 pseudodatasets with all systematic
uncertainties considered. We then measure the cross section in each of
the 2000 pseudodatasets at each input value and assess linearity. A
linear fit to the measured vs. input cross section gives a slope of
0.99$\pm$0.01 and intercept of $-0.14\pm0.05$. Therefore, over the
range considered, there is no significant evidence of bias in the
measurement procedure.

\begin{figure}
\centering
\includegraphics[width=0.45\textwidth]{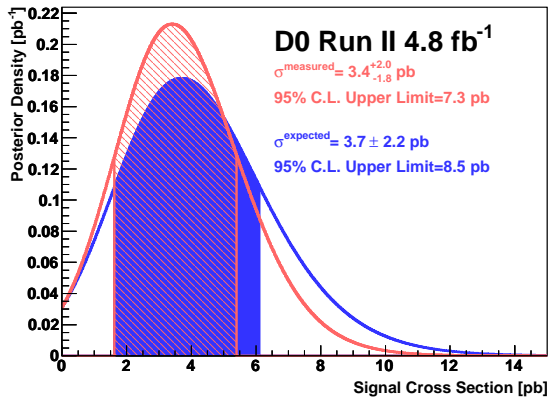}
\caption{Expected SM and measured Bayesian posterior probability
  densities for the $tb$+$tqb$ cross section. The shaded regions
  illustrate $\pm$ one standard deviation from the peak locations.
}\label{fig.post}
\end{figure}

We obtain an observed posterior density that is used to define an
upper limit on the cross section assuming no signal. We can use the
same technique to determine an observed cross section and its
uncertainty.  Assuming no signal, we extract an upper limit of
\TauLimObs~pb at 95\% C.L. If we perform a cross section measurement,
we obtain \TauXSobs~pb.  The measured sensitivity, a ratio of the
position of the peak of the measured posterior density to its lower
half width, is \TauXSobsPeakWidth.  Figure~\ref{fig.post} shows the
expected and measured posterior densities with shaded regions
corresponding to $\pm$ one standard deviation from the peak locations.

\section{$t\bar{t}$ Cross Check}
As an additional cross check of our background model, we have measured
the top quark pair production cross section in the same data sample,
including systematic uncertainties and using the same background model
and the same techniques as we use to measure the single top quark
cross section.  We measure a top quark pair production cross section
of \ttbarXSobs, in good agreement with the theoretical expectation,
7.91$^{+0.61}_{-0.56}$~pb, from the next-to-next-to-leading-order
calculation for a top quark mass of 170~GeV~\cite{ttbar-xsec}, and a
recent D0 experimental result, $8.18^{+0.98}_{-0.87}$~pb, also for the same
top quark mass~\cite{ttbar-xsec-exp}.

\section{Combination with Other Channels}\label{sect.combine}
\begin{figure}[h]
  \centering
  \includegraphics[width=0.45\textwidth]{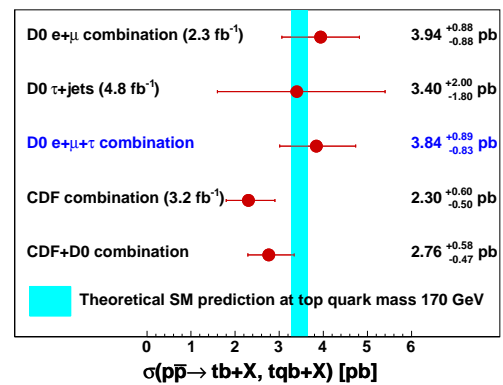}
\caption{Summary of several recent measurements of single top
  quark production cross section. The theoretical SM prediction~\cite{st-theory} at
  a top quark mass of 170~GeV is included as a shaded band. The ``D0
  e+$\mu$ combination'' result is taken from~\cite{d0_st_prl2009}
  while the ``CDF combination'' result comes
  from~\cite{cdf_st_prl2009} and the ``CDF+D0 combination'' result
  from~\cite{cdf_d0_st_2009}.  }\label{fig.allXS}
\end{figure}

As this data sample has no overlap with that used
in~\cite{d0_st_prl2009}, it is straightforward to combine the results.
In the combination, the tau channel and the (electron,muon)+jets
channels are treated as two independent channels using the same
Bayesian approach used to combine different tau channels above. The
ratio of the position of the peak of the expected posterior density to
its lower half width is \SuperCombineXSexpPeakWidth, compared to
\EMUCombineXSexpPeakWidth\ in the electron+jets and muon+jets channels
combined. We gain \SuperCombineExpGainInPeakWidth\ in expected
sensitivity by adding the tau+jets channel. The observed posterior
density is also calculated and yields a combined cross section of:
\begin{equation*}
  \sigma(p\bar{p}\rightarrow tb+X, tqb+X) = \SuperCombineXSall \mathrm{\ pb}
\end{equation*}
Figure~\ref{fig.allXS} shows several recent measurements of single top
quark production compared to the theoretical SM
prediction~\cite{st-theory}, 3.46$\pm$0.18~pb, calculated for a top
quark mass of 170~GeV~\cite{st-theory}.

\section{Summary}
In summary, we have presented the first direct study for single top
quark production in the tau+jets channel using 4.8~fb$^{-1}$ of
integrated luminosity at the D0 experiment. Due to different dominant
backgrounds and different systematic uncertainties from the electron
and muon channels, the tau+jets channel serves as a channel to
independently search for single top quarks. To increase sensitivity,
electron+jets events not entering the measurement in the electron+jets
channel and where the electron satisfies tau identification criteria
are also included in the tau+jets sample. An upper limit of
\TauLimObs~pb at the 95\% C.L.  for the cross section is obtained. The
expected sensitivity of the tau+jets channel alone is
\TauXSexpPeakWidth\ standard deviations.  Adding the tau+jets channel
increases the signal acceptance by 32\% compared to the D0 observation
analysis, which was based on electron+jets and muon+jets channels. The
expected sensitivity of the electron+jets, muon+jets and tau+jets
combined analysis is \SuperCombineXSexpPeakWidth\ standard deviations,
to be compared to \EMUCombineXSexpPeakWidth\ standard deviations in
electron+jets and muon+jets alone. The measured cross section in all
three combined channels is found to be $\SuperCombineXSall$~pb. This
is the most precise measurement to date of the single top quark
production cross section.
\appendix
\section{Acknowledgements}

\input{acknowledgement_paragraph_r2.tex}
%\appendix*
{\color{black}
\section{Appendix I: Tau Identification Variable
  Definitions}\label{append.tauvar}

\begin{description}
\item[$\boldsymbol{\delta\alpha}$: ]
%   $\sqrt{(\frac{\Delta\phi}{\sin{\theta}})^2+(\Delta
%       \eta)^2}$, where $\Delta\phi$ and $\Delta\eta$ are
%   differences between the vector sums of tau tracks and of all
%   EM~subclusters, $\theta$ is the azimuthal angle of the centroid of
%   the vector sum of EM~subclusters~\cite{paper.2005.D0.ZTauTau}.
  $\sqrt{(\Delta\phi)^2+(\Delta\eta)^2}$, where $\Delta\phi$ and $\Delta\eta$ are
  differences between the vector sums of tau tracks and of all EM~subclusters~\cite{paper.2005.D0.ZTauTau}.
\item[$\boldsymbol{e_{12}}$: ] $\sqrt{E_T^{\tau_{\mathrm{trk}}}\cdot E_T^{\mathrm{EMcls}}}$/$E^\tau_T$,
  where $E_T^{\tau_{\mathrm{trk}}}$ is the sum over all tau-associated
  tracks' $E_T$, $E_T^{\mathrm{EMcls}}$ is $E_T$ of the sum over
  EM~subclusters. For a system of tau-associated tracks and
  EM~subclusters, the observed tau transverse mass is $e_{12}\times E^\tau_T\times\delta\alpha$ in the small angle approximation~\cite{paper.2005.D0.ZTauTau}.
\item[$\boldsymbol{E_T^\mathrm{alltrk-trk1-trk2}}$: ] transverse energy of tracks except the first two
  highest-$p_T$ tracks.
\item[$\boldsymbol{E_T^\mathrm{EMcl2}}$: ] transverse energy of the second-highest-$p_T$ EM subcluster.
%\item[$E_T^\mathrm{EMcls}$: ] transverse energy of EM subclusters.
\item[$\boldsymbol{E_T^\mathrm{EMlayer3}}$:] transverse energy deposited in the 3$^\mathrm{rd}$ layer
  of the EM calorimeter within a cone $\mathcal{R}<0.5$.
\item[$\boldsymbol{E^\mathrm{trk1}_T}$: ] transverse energy of the highest-$p_T$ track.
\item[Isolation: ] $\frac{\sum p_T^{\mathrm{trk}}}{\sum
    p_T^{\tau_{\mathrm{trk}}}}$, where $\sum p_T^{\mathrm{trk}}$ is
  the sum of $p_T$ of non-tau-associated tracks within a cone size 0.5 and
  $\sum p_T^{\tau_{\mathrm{trk}}}$ is the sum over all tau-associated
  tracks' $p_T$.
\item[Isolation$^\prime$: ] if $|\eta_{\mathrm{det}}|\leqslant1.0$, where
  $\eta_\mathrm{det}$ is tau's detector pseudorapidity, which is defined with respect to the
  center of the detector, Isolation$^\prime=\mathrm{Isolation}$.
If $|\eta_{\mathrm{det}}|>1.0$,  $\mathrm{Isolation}^\prime=\mathrm{Isolation}/(1.5\times|\eta_{\mathrm{det}}|-0.5)$. 
\item[Profile: ] $\frac{E_{T1} + E_{T2}}{E_T^\tau}$ where $E_{T1}$
  and $E_{T2}$ are the transverse energies of the two highest-$p_T$
  calorimeter towers in a tau object.
\item[Profile$^\prime$: ] if $|\eta_{\mathrm{det}}|\leqslant$1.5, Profile$^\prime$=Profile.  If $|\eta_{\mathrm{det}}|>$1.5,
  Profile$^\prime=\mathrm{Profile}\times(0.67+0.22\times
  |\eta_{\mathrm{det}}|$). 
\item[Profile$_\mathrm{layer3}$: ] a ratio of $E_T$ of the highest $p_T$ EM~subcluster
  over $E_T$ deposited in the 3$^\mathrm{rd}$ layer of the EM calorimeter  within a cone $R<0.5$.
\item[Ratio$_\mathrm{EM12,\tau}$: ] $\frac{E^{\mathrm{EM1}}+E^{\mathrm{EM2}}}{E^\tau}$
  where $E^{\mathrm{EM1}}$ and $E^{\mathrm{EM2}}$ are energies
  deposited in the 1$^\mathrm{st}$ and 2$^\mathrm{nd}$ layers of the EM calorimeter.
\item[Ratio$_{\tau,\mathrm{trks}}$: ] $\frac{E^{\tau}_T}{E^{\tau}_T + \sum
    p_T^{{\tau}_\mathrm{trk}}}$
\item[Width$_{\eta,\phi}(\tau)$:] tau shower width, the root sum of squares of the $E_T$-weighted $\eta$-$\phi$ distance of all calorimeter towers with
  respect to the tau axis, i.e.,
  $\sqrt{\sum_{i=1}^n\left(\Delta\eta_i^2+\Delta\phi_i^2\right)\frac{E_{Ti}}{E_T}}$
  where $i$ is the index of calorimeter towers and
  $E_T=\sum_\mathrm{i}E_{Ti}$.
\item[Width$^\prime_{\eta,\phi}(\tau)$:] Width$_{\eta,\phi}(\tau)/(1.0+0.29\times|\eta_\mathrm{det}|)$.
\item[$\boldsymbol{z^\mathrm{trk1}_\mathrm{DCA}}$: ] $z$ position of the highest-$p_T$ track at DCA.
\end{description}
}

%%%%%%%%%%%%%%%%%%%%%%%%%%%%%%%%%%%%%%%%%%%%%%%%%%%%%%%%%%%%
% _     _ _     _ _                             _           
%| |__ (_) |__ | (_) ___   __ _ _ __ __ _ _ __ | |__  _   _ 
%| '_ \| | '_ \| | |/ _ \ / _` | '__/ _` | '_ \| '_ \| | | |
%| |_) | | |_) | | | (_) | (_| | | | (_| | |_) | | | | |_| |
%|_.__/|_|_.__/|_|_|\___/ \__, |_|  \__,_| .__/|_| |_|\__, |
%                         |___/          |_|          |___/ 
%-----------------------------------------------------------

%\end{multicols}
\end{document}

%% file: list_of_authors_r2.tex
% LIST_OF_AUTHORS_R2.TEX                 10/14/09           
%
\author{V.M.~Abazov$^{37}$}
\author{B.~Abbott$^{75}$}
\author{M.~Abolins$^{65}$}
\author{B.S.~Acharya$^{30}$}
\author{M.~Adams$^{51}$}
\author{T.~Adams$^{49}$}
\author{E.~Aguilo$^{6}$}
\author{M.~Ahsan$^{59}$}
\author{G.D.~Alexeev$^{37}$}
\author{G.~Alkhazov$^{41}$}
% ...

\affiliation{\vspace{0.1 in}(The D\O\ Collaboration)\vspace{0.1 in}}
% ...
\affiliation{$^{6}$University of Alberta, Edmonton, Alberta, Canada;
                Simon Fraser University, Burnaby, British Columbia, Canada;
                York University, Toronto, Ontario, Canada and
                McGill University, Montreal, Quebec, Canada}
% ...
\affiliation{$^{30}$Tata Institute of Fundamental Research, Mumbai, India}
% ...
\affiliation{$^{37}$Joint Institute for Nuclear Research, Dubna, Russia}
% ...
\affiliation{$^{41}$Petersburg Nuclear Physics Institute,
                St. Petersburg, Russia}
% ...
\affiliation{$^{49}$Florida State University, Tallahassee, Florida 32306, USA}
% ...
\affiliation{$^{51}$University of Illinois at Chicago,
                Chicago, Illinois 60607, USA}
% ...
\affiliation{$^{59}$Kansas State University, Manhattan, Kansas 66506, USA}
% ...
\affiliation{$^{65}$Michigan State University,
                East Lansing, Michigan 48824, USA}
% ...
\affiliation{$^{75}$University of Oklahoma, Norman, Oklahoma 73019, USA}
% ...

%% file: acknowledgement_paragraph_r2.tex
% acknowledgement_paragraph_r2.tex                         10/14/09
%
We thank the staffs at Fermilab and collaborating institutions, 
and acknowledge support from the 
DOE and NSF (USA);
CEA and CNRS/IN2P3 (France);
FASI, Rosatom and RFBR (Russia);
CNPq, FAPERJ, FAPESP and FUNDUNESP (Brazil);
DAE and DST (India);
Colciencias (Colombia);
CONACyT (Mexico);
KRF and KOSEF (Korea);
CONICET and UBACyT (Argentina);
FOM (The Netherlands);
STFC and the Royal Society (United Kingdom);
MSMT and GACR (Czech Republic);
CRC Program, CFI, NSERC and WestGrid Project (Canada);
BMBF and DFG (Germany);
SFI (Ireland);
The Swedish Research Council (Sweden);
and
CAS and CNSF (China).
%